\documentclass{nature-Fig}

\pdfoutput=1

\usepackage{amsmath}
\usepackage{amssymb}
\usepackage{amsfonts}
\usepackage{graphicx}

\newcommand{\K}[1]{$|#1\rangle$}

\title{Coherent quantum state storage and transfer between two phase qubits via a resonant cavity}

\author{Mika A. Sillanp\"a\"a$^1$, Jae I. Park$^1$, Raymond W. Simmonds$^1$}

\begin{document}
\maketitle
\begin{affiliations}
\item National Institute of Standards and Technology, 325
Broadway, Boulder CO 80305, USA
\end{affiliations}

\begin{abstract}
A network of quantum-mechanical systems showing long lived phase
coherence of its quantum states could be used for processing
quantum information. As with classical information processing, a
quantum processor requires information bits (qubits) that can be
independently addressed and read out, long-term memory elements to
store arbitrary quantum states\cite{memory04,trap05}, and the
ability to transfer quantum information through a coherent
communication bus accessible to a large number of
qubits\cite{Falci03,cleland04}. Superconducting qubits made with
scalable microfabrication techniques are a promising candidate for
the realization of a large scale quantum information
processor\cite{Bouchiat98,Nakamura99,CPqubit,FluxQBExp,MSS}.
Although these systems have successfully passed tests of coherent
coupling for up to four
qubits\cite{tsai2QB,CoupledPQ,Steffen06,ilichev06}, communication
of individual quantum states between qubits via a quantum bus has
not yet been demonstrated. Here, we perform an experiment
demonstrating the ability to coherently transfer quantum states
between two superconducting Josephson phase qubits through a
rudimentary quantum bus formed by a single, on chip,
superconducting transmission line resonant cavity of length 7 mm.
After preparing an initial quantum state with the first qubit,
this quantum information is transferred and stored as a
nonclassical photon state of the resonant cavity, then retrieved
at a later time by the second qubit connected to the opposite end
of the cavity. Beyond simple communication, these results suggest
that a high quality factor superconducting cavity could also
function as a long term memory element. The basic architecture
presented here is scalable, offering the possibility for the
coherent communication between a large number of superconducting
qubits.
\end{abstract}


A particularly interesting quantum information architecture
involves the interaction of matter and quantized electromagnetic
fields, or cavity Quantum Electro Dynamics (QED). In some cavity
QED systems, atoms, which can play the role of qubits, are passed
through or are trapped within optical or microwave cavities with
resonant modes matching one of the atom's spectral lines. These
systems\cite{HarocheRaimond,SchleichWalther} have enabled
fundamental tests of quantum mechanics, as well as demonstrations
of quantum memory and a quantum bus\cite{OpticsSwap97}. Recently,
the Cooper pair box\cite{Bouchiat98} has been successfully
incorporated into a superconducting resonant cavity in order to
perform analogous experiments in the strong coupling regime,
forming a new field known as ``circuit
QED"\cite{buisson01,Blais04,CQED,chiorescu04,johansson06}. Similar
resonant cavities have also been used to stabilize flux
qubits\cite{KochPRL06}, and thus far experiments have found
spectroscopic evidence for the entanglement between two phase
qubits and a resonator\cite{WellstoodPRL05}. In this work, we
report the first time-domain measurements showing coherent
interactions for circuit QED performed using superconducting
Josephson phase qubits coupled to a cavity formed by a
transmission-line resonator. Moreover, by coupling two phase
qubits to a single cavity, taking advantage of the independent
control of each phase qubit and single-shot readout, we have
constructed an elementary quantum memory and quantum bus in a
superconducting system.

\begin{figure}[!p] 
\linespread{1.5}
\center
\includegraphics[width=14cm]{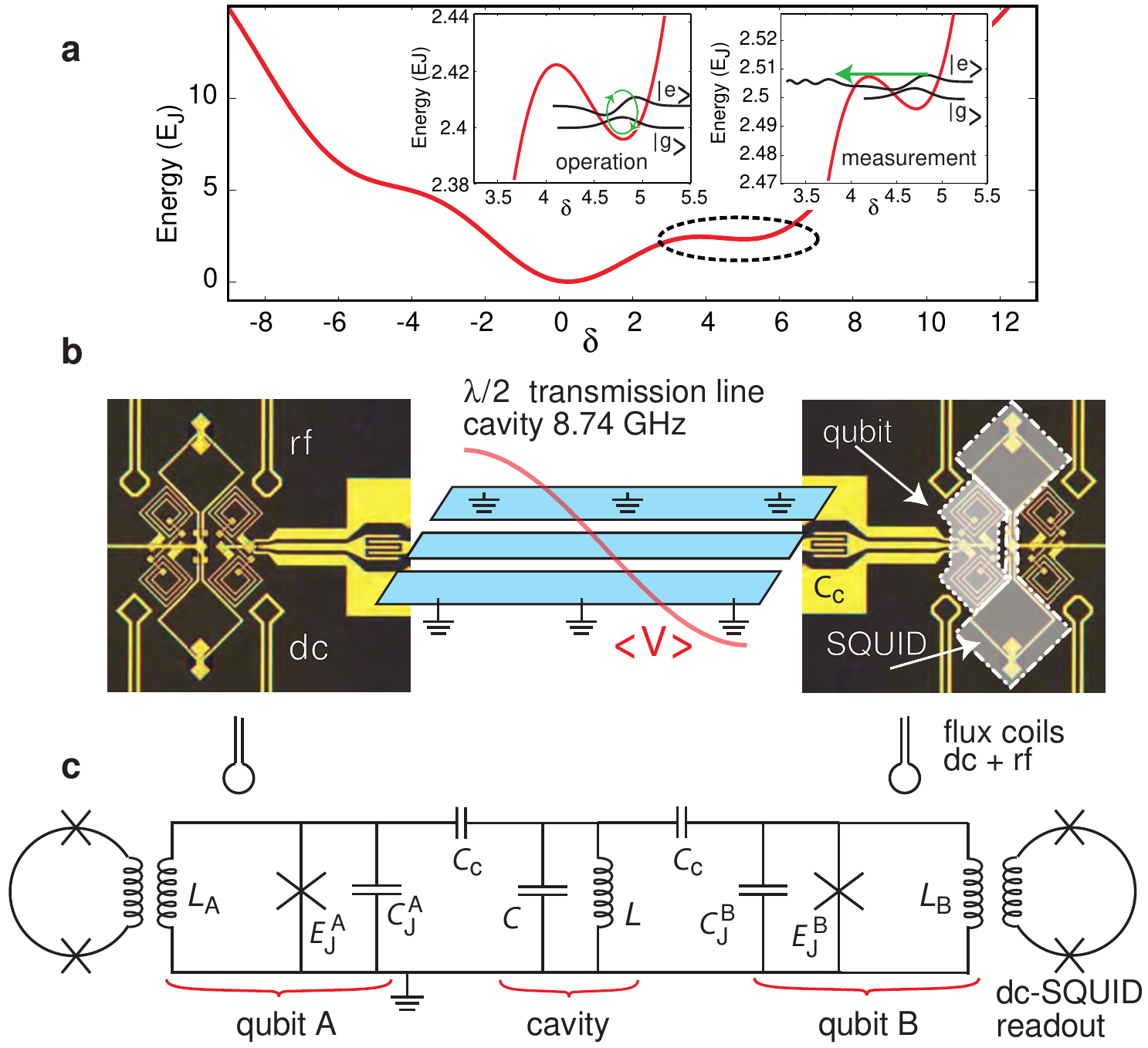}
\caption{\textbf{a}, Potential energy diagram of the phase qubit
and illustration of the measurement, where tunneling of the qubit
excited state \K{e} results in a difference of about one flux
quantum in the loop, which is read out by a dc
SQUID\cite{SimmondsPRL04}. We determine the excited state
population $P_A$ and $P_B$ of qubit $A$ and $B$ by repeating
simultaneous single shot measurements\cite{CoupledPQ} thousands of
times. \textbf{b}, Illustration of the quantum memory element with
two Josephson phase qubits (with junction areas \mbox{$\simeq 14$
$\mu$m$^2$}) connected via coupling capacitors $C_c \simeq 6.2$ fF
to either end of a resonant cavity formed by a \mbox{7 mm} long
slowly meandering coplanar waveguide, with the qubits separated by
about \mbox{1.1 mm}. The red line depicts the voltage amplitude of
the lowest \mbox{$\lambda/2$-mode}. The device was fabricated with
standard optical lithography, producing
\mbox{Al/AlO$_\mathrm{x}$/Al} junctions on a sapphire substrate,
using SiN$_\mathrm{x}$ as an insulator between the metallic
layers. \textbf{c}, Lumped element equivalent circuit near the
$\lambda/2$ resonance. The cavity has an effective inductance
\mbox{$L\simeq 580$ pH} and capacitance \mbox{$C\simeq 0.57$ pF},
and both qubits had roughly \mbox{$L_{A,B}\simeq 690$ nH},
\mbox{$E_J^{A,B}\simeq 45$ K}, \mbox{$C_J^{A,B}\simeq 0.7$ pF}.}
\label{2xQB16CoupSchema} 
\end{figure}

For a flux-biased Josephson phase qubit\cite{SimmondsPRL04}, the
ground state \K{g} and the first excited state \K{e} are encoded
in the phase difference $\delta$ across a large-capacitance
superconducting Josephson junction placed in a superconducting
loop (Fig.~\ref{2xQB16CoupSchema}a). These states resemble those
of a simple harmonic oscillator but for the nonlinear, anharmonic
potential\cite{martinisqb} formed by the combination of the
Josephson coupling energy $-E_J\cos(\delta)$ and the inductive
energy stored in the superconducting loop, where $E_J$ is the
Josephson energy. Due to their large capacitance, addressability,
single-shot readout, and the ease with which the energy level
separation $\hbar\omega\equiv E_e - E_g$ can be tuned, phase
qubits have proven to be relatively easy to couple
together\cite{WellstoodPRL05,CoupledPQ}. Ultimately, most
superconducting qubit strategies have the ability to be connected
in various ways allowing for the possible formation of a quantum
processor consisting of both qubits and a set of communication
channels or a ``qubus".

Our superconducting quantum system is presented in
\mbox{Fig.~\ref{2xQB16CoupSchema}b,c}. Both qubits $A$ and $B$ are
inductively coupled to two separate flux bias coils: one set of
coils is used to adjust a static, dc flux bias, whereas the other
set of rf coils, with a bandwidth from dc up to about 20 GHz,
enables rapid flux bias changes (``shift pulses"), inductively
coupled microwave pulses, and a fast measurement pulse. Each set
of qubit dc flux bias lines includes low-pass and copper powder
filters, while each set of rf flux pulsed lines are combined into
a single microwave coaxial line at room temperature and attenuated
by roughly 40 dB inside the cryostat. Microwave pulse control is
performed with passively filtered (roughly gaussian shaped pulses)
and standard microwave mixers. Independently addressable state
readout is accomplished via inductively coupled dc superconducting
quantum interference devices (SQUIDs).

Our resonant cavity is an open ended coplanar waveguide whose
lowest standing wave eigenmode ($\lambda/2$-mode) has voltage
maxima at each end of the waveguide
(\mbox{Fig.~\ref{2xQB16CoupSchema}b}). Near resonance this
waveguide acts like a parallel, lumped element resonant circuit
(\mbox{Fig.~\ref{2xQB16CoupSchema}c}). The $\lambda/2$-mode forms
a simple harmonic oscillator with an energy $\hat{H}_r =
\hbar\omega_{r}(\hat{a}^\dagger \hat{a} + \frac{1}{2})$ at the
frequency $\omega_{r}/2\pi = 1/2\pi\sqrt{LC} \simeq 8.74$ GHz,
where $L = 2Z_0/\pi\omega_r$ and $C = \pi/2\omega_rZ_0$ represent
their lumped element equivalents\cite{Pozar}, $Z_0\sim 50$
$\Omega$ is the characteristic impedance of the coplanar
waveguide, and the raising and lowering operators
$\hat{a}^\dagger$ and $\hat{a}$ increase or decrease the photon
number in the cavity.

The hamiltonian of our quantum system formed by a single resonant
cavity coupled at both ends to qubits $A$ and $B$, respectively,
has the form of the Jaynes-Cummings hamiltonian familiar from
quantum optics:
\begin{equation}
\hat{H} = \hat{H}_r + \sum_{j=A,B}\hat{H}_j + \sum_{j=A,B}\hbar
g_j(\hat{a}^\dagger\hat{\sigma}_-^j + \hat{a}\hat{\sigma}_+^j)
\end{equation}
where $\hat{H}_j ={1\over2}
\hbar\omega_j\hat{\sigma}_+^j\hat{\sigma}_-^j$ is the single qubit
hamiltonian, $\hat{\sigma}_+^j$ ($\hat{\sigma}_-^j$) is the
raising (lowering) operator for creating (annihilating)
excitations in the $j$th-qubit, and $\hbar\omega_j$ is controlled
by the amplitude of the dc and rf flux bias. The interaction
energy, $2g_{A,B}\sim\omega_r(C_c/\sqrt{CC_J^{A,B}})$, was
designed to be large enough to ensure that the time scale of
quantum state transfer, \mbox{$\pi/g_{A,B}\sim 10$ ns}, would not
be limited by the relaxation times of either qubit or the cavity,
putting this experiment in the strong coupling regime
\mbox{($g_{A,B}>\gamma_{A,B}>\kappa$)} for circuit QED\cite{CQED}
with qubit decay rates of \mbox{$\gamma_{A,B}\sim 5-20$ MHz}, and
a cavity decay rate of \mbox{$\kappa/2\pi\lesssim 1$ MHz}.

When a single qubit is on resonance with the cavity, so that the
detuning is $\Delta\equiv\omega - \omega_r = 0$, the individual
eigenstates of the qubit (\K{g},\K{e}) and the cavity
(\K{0},\K{1}) are no longer the eigenstates of the coupled system.
Here, we find new eigenstates formed by an equal combination of
cavity and qubit photons, leading to the symmetric and
antisymmetric superpositions, (\K{0}\K{e} $\pm$
\K{1}\K{g})$/\sqrt{2}$. We also find that the energy level
separation of the new eigenstates, $\hbar(\omega\pm g)$, shows the
typical vacuum Rabi mode splitting.

In addition, the exchange of photons between the cavity and a
single qubit is strongest on resonance. In a familiar cavity
QED-process, a single off-resonant atom or qubit is excited,
\K{e}, and is then rapidly brought into resonance with the empty
cavity, \K{0}. Here the initial coupled-system state \K{0}\K{e}
begins to oscillate in time according to $\cos(gt)$\K{0}\K{e}$ -
i\sin(gt)$\K{1}\K{g}, so that the qubit photon, \K{e}, is
transformed into a cavity photon, \K{1}, after a time $t = \pi/2g$
set by the interaction energy $\hbar g$. This process continues
coherently with the photon continuously being transferred back and
forth between the qubit and the cavity in what is known as vacuum
Rabi oscillations. Our phase qubits play the role of atoms in the
analogous quantum optical system in which the interaction time $t$
is controlled by the atom's velocity through the cavity, while in
our system, we have the flexibility of using fast (\mbox{$\sim 1$
ns} rise time), roughly rectangular flux bias shift pulses with
adjustable amplitude (detuning $\Delta$) and width (interaction
time $t$).

\begin{figure}[!p] 
\linespread{1.5} \center
\includegraphics[width=16cm]{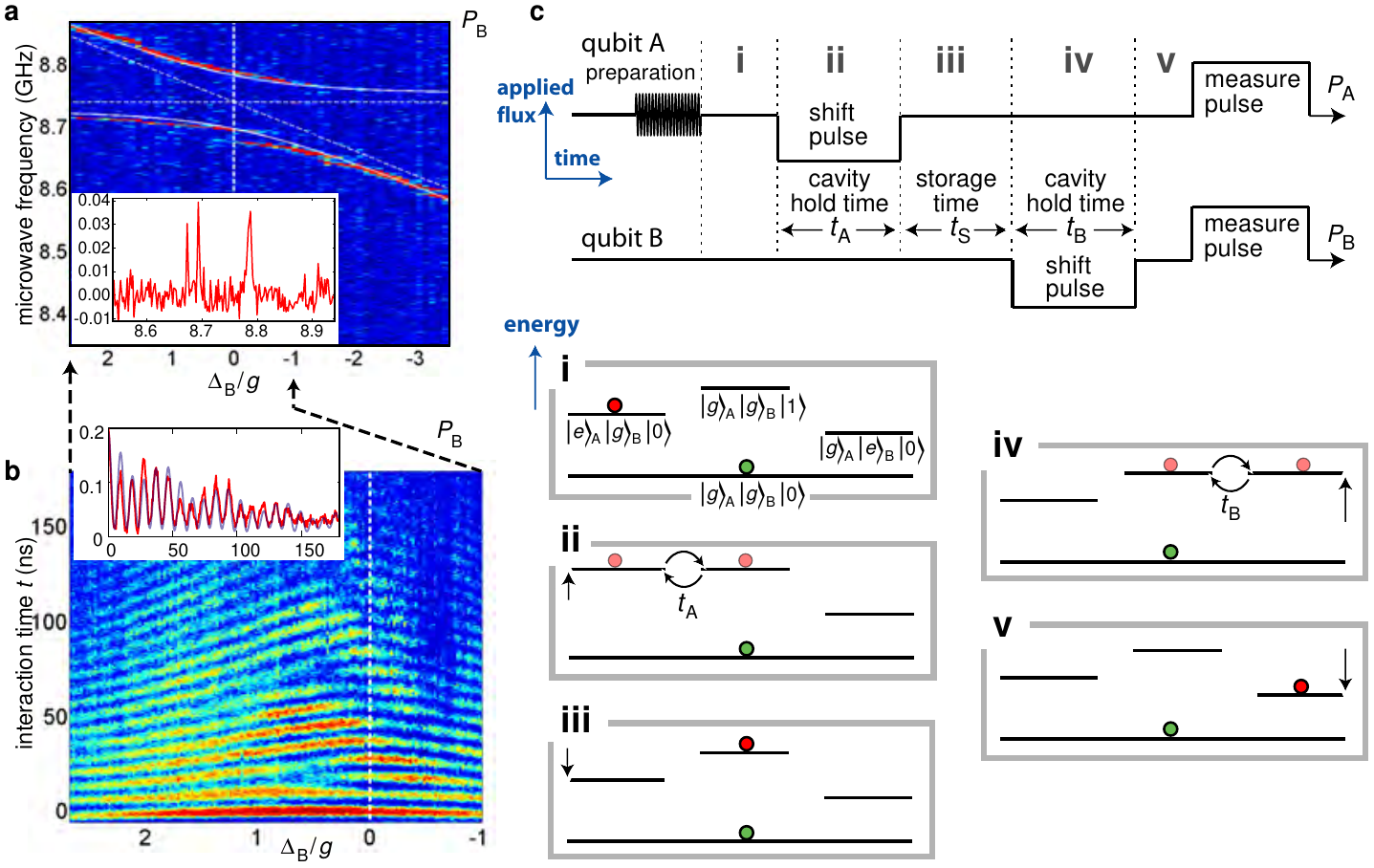}
\caption{\textbf{a}, Microwave spectroscopy of qubit $B$ showing
the vacuum Rabi splitting (qubit $A$ detuned). Blue color
represents low $P_B$, red represents high $P_B$. The inset shows a
cross-section at $\Delta_B = 0$ (along the dashed line).
\textbf{b}, Vacuum Rabi oscillations in qubit $B$ after a short
\mbox{$\pi$ pulse}. The inset shows a cross-section at $\Delta_B =
0$. The blue line shows the numerical results including the TLS
defect near resonance. \textbf{c}, Illustration of the general
quantum state transfer protocol performed by a sequence of flux
bias pulses applied to qubit $A$ and $B$. Here each qubit is
effectively decoupled from the cavity, except during the shift
pulses, which bring them into resonance with the cavity, one qubit
at a time. \textbf{i}, An arbitrary superposition state
$\alpha$\K{g}$_A$ + $\beta$\K{e}$_A$ is prepared in qubit $A$. The
red and green shaded circles represent mixtures of the occupied
energy levels. \textbf{ii}, Qubit $A$ is shifted into resonance
with the cavity for an interaction time lasting one half of a
vacuum Rabi period, $t_A = \pi/2g$, the photon has been exchanged
and the state of qubit $A$ has been mapped into a superposition
\mbox{$\alpha$\K{0} + $\beta$\K{1}} of the two lowest photon
number eigenstates (Fock states) of the cavity. \textbf{iii},
Qubit $A$ is shifted off resonance, storing the initial state in
the cavity for a time duration $t_S$. \textbf{iv}, Qubit $B$ is
shifted into resonance for one half of a vacuum Rabi period, $t_B
= \pi/2g$, transferring the state into qubit $B$, leaving the
cavity in its ground state \K{0}. \textbf{v}, Both qubits are
detuned, completing the coherent quantum state transfer from qubit
$A$ to qubit $B$.} \label{2xQB16CoupPulses}
\end{figure}

As a first demonstration of strongly coupled circuit QED in our
system, these two basic vacuum Rabi behaviors were independently
verified for each qubit, $A$ and $B$. In
\mbox{Fig.~\ref{2xQB16CoupPulses}a}, we show an example of the
vacuum Rabi splitting for qubit $B$ (a similar splitting was
obtained for qubit $A$) using well established spectroscopic
techniques\cite{SimmondsPRL04,simmonds04}. Vacuum Rabi
oscillations were also obtained for both qubits using an analogous
technique borrowed from quantum optics\cite{brune96} and utilized
previously\cite{simmonds04} for a superconducting flux qubit
coupled to a lumped-element cavity\cite{johansson06}. With qubit
$B$ fixed at a given detuning $\Delta_B$, a fast (\mbox{$\sim 4$
ns}) \mbox{$\pi$ pulse} was applied to the qubit inducing vacuum
Rabi oscillations with a raw contrast of \mbox{$\sim 20$ \%},
visible out to \mbox{200 ns}. In
\mbox{Fig.~\ref{2xQB16CoupPulses}b}, we show an example of vacuum
Rabi oscillations for qubit $B$ (similar oscillations were
obtained for qubit $A$) for various detunings $\Delta_B$. We see
an increase in the vacuum Rabi frequency with detuning, roughly as
$\sqrt{4g_B^2+\Delta_B^2}$, with a minimum value on resonance
($\Delta_B=0$). An additional energy splitting, near the cavity
resonance (seen in \mbox{Fig.~\ref{2xQB16CoupPulses}a} on the
lower spectroscopic branch) caused by a two-level system (TLS)
defect common to large area Josephson phase
qubits\cite{SimmondsPRL04,simmonds04}, is responsible for a slight
broadening of the spectroscopic splitting and a beating in the
oscillations centered at $\Delta_B/g_B\sim 0.5$. Numerical
calculations taking into account the size and position of the TLS
agree well with the data for $g_B/\pi\approx 86$ MHz, where a
small amount of beating is still visible on resonance (see the
inset of \mbox{Fig.~\ref{2xQB16CoupPulses}b}). Both qubits showed
similar behavior (without a nearby TLS in qubit $A$), different by
less than \mbox{10 \%}, with a coupling strength ($g_A\approx
g_B$) matching the design values (see
Fig.~\ref{2xQB16CoupSchema}). After calibrating the amplitude of
the shift pulses separately, for both qubits at their far-detuned
operation points, we remeasured the vacuum Rabi oscillations using
the shift pulse sequence described in
\mbox{Fig.~\ref{2xQB16CoupPulses}c}. Both experimental methods
gave similar results with a reduced contrast due to nonoptimized
shift pulse shaping and induced Landau-Zener transitions between
the qubit and distributed TLS defects\cite{simmonds04}.

In order to investigate the transfer of quantum states through the
resonant cavity, we utilize the vacuum Rabi interaction of both
qubits. The complete sequence (i-v) is described in
\mbox{Fig.~\ref{2xQB16CoupPulses}c}. Using the static, dc flux
bias coils the phase qubits are completely detuned
($\Delta_{A,B}\sim 15g_{A,B}$) from the cavity and each other to
suppress any stray cavity and qubit interactions. In this
configuration, we first, (i) prepare a superposition state for
qubit $A$ using a rapid microwave pulse. Next, (ii) we apply a
shift pulse to qubit $A$, placing it on resonance with the cavity
for a time duration $t_A$. With shift pulse speeds much greater
than $g_A/2\pi$ but still much less than $\omega_A/2\pi$ (still
adiabatic), we effectively preserve the initially prepared quantum
state until $\Delta_A = 0$, when the vacuum Rabi oscillations
begin to mix the qubit-cavity states. (iii) With the detuning of
qubit $A$ restored, we wait for a short storage time
\mbox{$t_S\lesssim 10$ ns} before, (iv) a second shift pulse
places qubit $B$ on resonance with the cavity for a time $t_B$.
Finally, (v) qubit $B$ is returned to its fully detuned position
and both qubits are measured simultaneously using a fast
(\mbox{$\sim 4$ ns}) flux bias measurement
pulse\cite{CoupledPQ,simmonds04} that reveals the excited state
occupation probabilities $P_A$ and $P_B$ corresponding to qubits
$A$ and $B$, respectively.

\begin{figure}[!p] 
\center
\includegraphics[width=16cm]{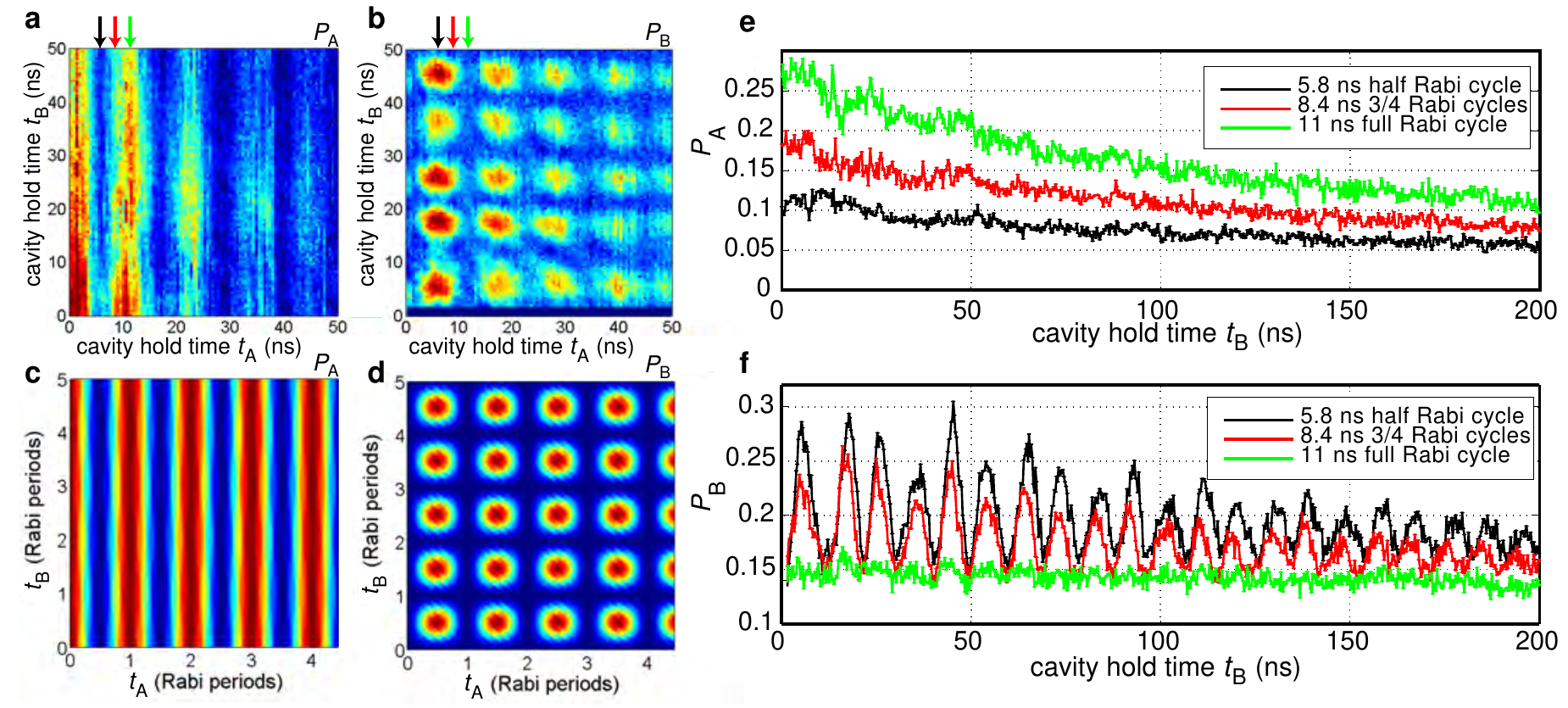}
\caption{Experimental data showing the quantum state transfer from
qubit $A$ to $B$ via the cavity, according to the protocol in
Fig.~\ref{2xQB16CoupPulses}, where the qubit $A$ excited state
\K{e}$_A$ is first mapped into the single photon state \K{1} in
the cavity, and then transferred into qubit $B$.
\textbf{a},\textbf{b}, Measured populations of qubits $A$ and $B$
as functions of the cavity hold times. Blue color represents low
$P_{A,B}$, red represents high $P_{A,B}$. \textbf{c}, \textbf{d},
Corresponding theoretical prediction for ideal conditions without
decoherence and 100 \% fidelity. \textbf{e}, Excited state
occupancy $P_A$ of the source qubit $A$ reveals a lower population
if the interaction time equals one half of a vacuum Rabi period,
\mbox{$t_A = \pi/2g\sim 5.8$ ns} (black). \textbf{f}, Simultaneous
measurement of qubit $B$ shows vacuum Rabi oscillations induced by
the transfer of a single photon (black).  Here the black, red, and
green curves in \textbf{e} (\textbf{f}) correspond to data
indicated by the arrows in \textbf{a} (\textbf{b}). For a full
discussion see the text.} \label{2xQB16CoupData}
\end{figure}

For the the experimental data shown in
\mbox{Fig.~\ref{2xQB16CoupData}}, we used the state transfer
protocol as outlined in \mbox{Fig.~\ref{2xQB16CoupPulses}c} with
an initial microwave $\pi$-pulse applied to qubit $A$ to create a
simple pure state \K{e}$_A$ for transfer.
\mbox{Fig.~\ref{2xQB16CoupData}a,b} show data over a range of
interaction times $t_A$ and $t_B$. The population maxima (color
scale) in the target qubit $B$ in
\mbox{Fig.~\ref{2xQB16CoupData}b} satisfy the following
conditions: whenever $t_A$ is an odd half-multiple of a vacuum
Rabi period, qubit $A$ has a low population $P_A$ and we see a
corresponding vacuum Rabi oscillation of $P_B$ occurring in qubit
$B$. The experimental data is in good agreement with theoretical
calculations of equation~(1) under ideal conditions,
\mbox{Fig.~\ref{2xQB16CoupData}c,d}.

For clarity, we have extracted a set of three curves from the
color plots of \mbox{Fig.~\ref{2xQB16CoupData}a,b} (arrows) and
displayed them in \mbox{Fig.~\ref{2xQB16CoupData}e,f}. If both
shift pulses last for a half vacuum Rabi period $\pi/2g_{A,B}$,
then the qubit photon is completely transferred into the cavity
and the subsequent excited state population $P_A$ is low, while in
the target qubit $B$, we simultaneously observe clear vacuum Rabi
oscillations (black curve). The fact that the oscillations start
from a minimum indicates the presence of a photon in the cavity at
the moment of state transfer to qubit $B$, as expected. Thus, the
photon must leave qubit $A$, enter the cavity, where it is stored
for a short time, and then be finally deposited in qubit $B$.
Repeating this experiment for a full vacuum Rabi period
(\mbox{$t_A = \pi/g_A \sim 11.6$ ns}, green curves) shows no
oscillations in $P_B$, also as expected, since the photon was
fully returned to qubit $A$ (as indicated by higher values of
$P_A$), leaving the cavity empty. The red lines illustrate an
intermediate case, with $t_A \sim 3/4$ of a vacuum Rabi period
yielding oscillations of lower amplitude but the same frequency.
Thus, we conclude that we can clearly transfer photons between two
phase qubits, through the resonant cavity, as well as store this
quantum information for a short time. Because superconducting
cavities tend to be more coherent than state-of-the-art qubits,
due to their simplicity, well defined, well separated, and fixed
resonant modes, extremely high quality factor superconducting
microwave resonators\cite{JPLNature03} may provide us with a
feasible long-term memory element for superconducting quantum
information systems.

\begin{figure}[!p] 
\center
\includegraphics[width=16cm]{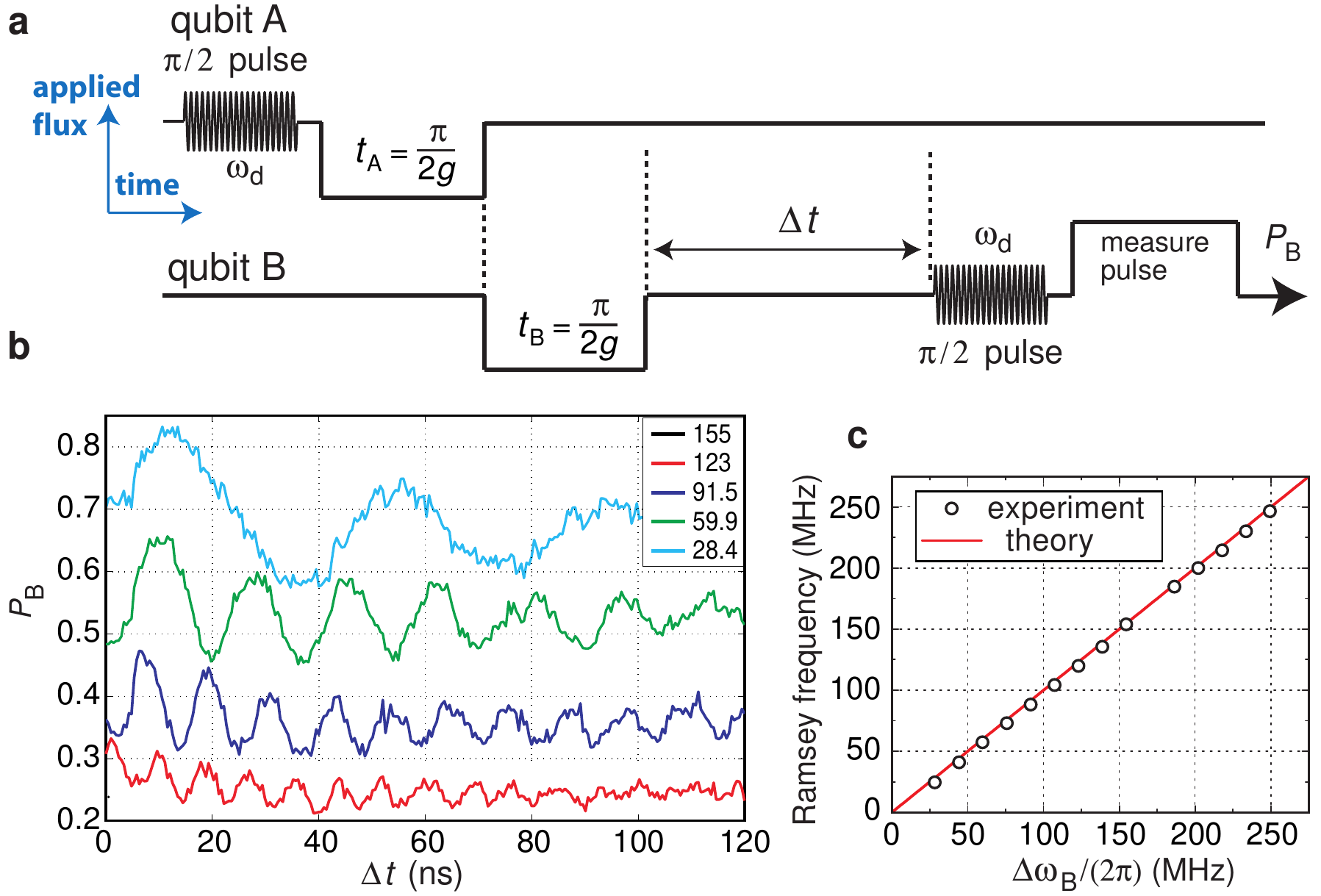}
\caption{Demonstration of the coherent transfer of a quantum state
through the quantum bus. We use $\pi/2$ microwave pulses detuned
from the level spacing frequencies $\omega_{A,B}$ of qubits $A$
and $B$ in order to perform a Ramsey fringe-type interference
experiment. \textbf{a}, We prepare an equal weight superposition
state (\K{g}$_A + $\K{e}$_A$)$/\sqrt{2}$ in qubit $A$ using a
\mbox{10 ns} long \mbox{$\pi/2$ pulse} (with frequency
$\omega_{d}$) while both qubits are detuned from the cavity and
from each other. We transfer this state into qubit $B$ as in
Fig.~\ref{2xQB16CoupPulses}, and then wait for a delay time
$\Delta t$ before applying a detuned \mbox{$\pi/2$ pulse} to qubit
$B$. This is analogous to Ramsey fringe experiments with single
qubits, where a coherent quantum state slowly precesses at the
microwave detuning frequency
$\Delta\omega_B\equiv\omega_B-\omega_d$. \textbf{b}, Coherent
oscillations in qubit $B$ for several detunings (vertically
displaced for clarity). \textbf{c}, The frequency of the
Ramsey-type oscillations as a function of the microwave detuning.
The solid line represents the theoretical predictions with no
fitting parameters.} \label{2xQB16CoupRamsey}
\end{figure}

In order to verify that quantum coherence is maintained during
state transfer for an arbitrary superposition state, we perform a
Ramsey fringe-type interference experiment\cite{CPqubit} that
preserves the quantum state up to a relative phase factor. We
follow a protocol (\mbox{Fig.~\ref{2xQB16CoupRamsey}a}) similar to
that used previously, except here, we first prepare qubit $A$ in
an equal-weight superposition state (\K{g}$_A +
$\K{e}$_A$)$/\sqrt{2}$, using a \mbox{$\pi/2$ pulse} applied
slightly off-resonance, $\Delta\omega_A\equiv\omega_d-\omega_{A}$,
where $\omega_d$ is the microwave drive frequency. Again, we
perform shift pulses in order, first, to map the initial state
onto a superposition of the two lowest photon number states \K{0}
and \K{1} of the cavity and, second, to retrieve this quantum
information through the transfer to the states \K{g}$_B$ and
\K{e}$_B$ spanned by qubit $B$. Following the coherent state
transfer to qubit $B$, we expect a clear precession of the
transferred state, (\K{g}$_B +
\exp(i\Theta)$\K{e}$_B$)$/\sqrt{2}$, during the time delay $\Delta
t$, where we have accumulated a relative phase shift $\Theta$
during the transfer process. By applying a final \mbox{$\pi/2$
pulse} to qubit $B$ (also slightly off-resonance, $\Delta\omega_B
= \Delta\omega_A$), we complete the Ramsey fringe-type experiment,
rotating qubit $B$ into a different state depending on the total
relative phase shift accumulated over the time $\Delta t$. In
\mbox{Fig.~\ref{2xQB16CoupRamsey}b,c}, we show the expected
Ramsey-type oscillations with frequencies linearly proportional to
the microwave detuning $\Delta\omega_B$, thus verifying the
transfer of quantum coherence through the cavity qubus.

In order to test the integrity of our experimental design, we
investigated in detail the possible role of stray unintended
photon generation in the cavity, both dc and rf inductive flux
cross-coupling between the two qubits, the role of nearby TLS
defects, and measurement cross-talk\cite{CoupledPQ} directly
through the cavity. First, we verified that the experiment
satisfied basic consistency checks based on predictions of the
model hamiltonian, equation~(1), by altering the transfer pulse
sequence shown in \mbox{Fig.~\ref{2xQB16CoupPulses}c}. When we
applied \mbox{$\pi$ pulses} to either qubit with any of the shift
pulses omitted, we saw no visible oscillations (above 1 \%
contrast) in the target qubit. When compared to the $\sim 20$ \%
contrast of the full state transfer sequence, this corresponds to
less than 0.05 stray photons in the cavity per \mbox{$\pi$ pulse}.
Next, we determined the cross coupling of shift pulses by studying
the flux modulation of one qubit for flux applied to the other
qubit. We found a leakage ratio of at most \mbox{6 \%} between the
two qubits, allowing us to avoid bias pulse cross-talk for large
detunings. We performed numerical simulations that included the
finite coherence times of each qubit, nearby TLS defects and no
additional cross talk. These results agree with the data as shown
in Fig.~\ref{2xQB16CoupPulses}. Finally, we performed detailed
time delay measurements\cite{CoupledPQ} in order to investigate
the role of measurement cross-talk when qubits $A$ and $B$ were
not measured simultaneously. These results show that using shift
pulses, which allows both qubits to be far detuned during
measurement, and the resonant cavity significantly reduce
measurement cross-talk. In this cavity-coupled phase qubit system,
the resonator between the qubits acts like an extremely narrow
bandpass filter (centered at its resonant frequency) which helps
block either qubit from the broadband transient microwave
excitations generated by the measurement process\cite{CoupledPQ}.

We have successfully coupled two superconducting Josephson phase
qubits through a resonant microwave cavity and have observed
vacuum Rabi splittings, vacuum Rabi oscillations, and the coherent
transfer and storage of quantum states mediated by the cavity. We
estimate that the fidelity of the state transfer protocol is
mostly limited by the quality of the phase qubits, the presence of
TLS defects, and the nonoptimization of the shape of the shift
pulses performing the state transfer. It is clear that further
measurements involving full state tomography$^{10}$ with higher
quality qubits must be performed in order to fully quantify the
fidelity of this cavity qubus. This simple demonstration, however,
clearly shows progress towards the storage and communication of
quantum information using coherent superconducting systems of
multiple qubits, an exciting new frontier for solid state circuit
QED and quantum information science.

\begin{addendum}
\item We gratefully acknowledge fruitful discussions with \mbox{J.
Aumentado}, \mbox{K. Cicak}, \mbox{K.\ Osborne}, \mbox{R.
Schoelkopf} and \mbox{David Wineland}. This work was financially
supported by NIST and DTO under grant number W911NF-05-R-0009.
Contribution of the U.S. government, not subject to copyright.
\end{addendum}

\begin{addendum}
\item[Competing interests statement] The authors declare that they
have no competing financial interests.
\end{addendum}

\begin{addendum}
\item[Correspondence] and requests for materials should be
addressed to simmonds@boulder.nist.gov
\end{addendum}

\bibliography{qed}

\end{document}